\documentclass[aps,twocolumn,pra,superscriptaddress,amsmath,showpacs,tightenlines]{revtex4-1}
\usepackage{bbm}
\usepackage{}
\usepackage{amssymb}
\usepackage{amsmath}
\usepackage{graphicx}
\usepackage{subfigure}
\usepackage{natbib}
\usepackage{epsfig}
\usepackage{amsfonts}
\usepackage{mathrsfs}
\usepackage{xcolor}
\usepackage[toc,page,title,titletoc,header]{appendix}

\begin{document}

\title{Quantum interferometry for rotation sensing in an optical  microresonator}
\author{Weijun Cheng}
\affiliation{Center for Quantum Sciences and School of Physics, Northeast Normal University,
Changchun 130024, China}
\author{Zhihai Wang}
\email{wangzh761@nenu.edu.cn}
\affiliation{Center for Quantum Sciences and School of Physics, Northeast Normal University,
Changchun 130024, China}
\author{Xiaoguang Wang}
\email{xgwang1208@zju.edu.cn}
\affiliation{Zhejiang Institute of Modern Physics, Department of Physics, Zhejiang University,
Hangzhou 310027, China}

\begin{abstract}
We theoretically propose a scheme to perform rotation sensing in a whispering-gallery-mode resonator setup. With the assistance of a large detuned two-level atom, which induces the effective coupling between clockwise and counterclockwise propagating modes in the resonator, we realize an effective interferometry with SU(2) algebraic structure. By studying the quantum Fisher information of the system, we find that the estimate accuracy for the angular velocity of the rotation can achieve and even break the Heisenberg limit in linear and nonlinear setup, respectively. The high performance of quantum metrology is proved to be associated with the state compressibility during the time evolution. We hope that our investigation will be useful in the design of a quantum gyroscope based on spinning resonators.
\end{abstract}

\maketitle

\section{introduction}
The ultraprecise estimation of parameters, which is introduced in the quantum interferometry~\cite{Caves1,Caves2}, has been widely promoted to the optical  microresonator. The theory claims that the quantum features such as entanglement and squeeze can dramatically enhance the interferometer sensitivity~\cite{V. Giovannetti1,V. Giovannetti2,D. W. Berry,J. Joo}. In this community, the phase sensitivity of interferometer can approach the Heisenberg limit with $\Delta x\sim 1/n$ ($n$ is the amount of the source employed) scaling, which is much better than the classical astrict, named the standard quantum limit with $\Delta x\sim 1/\sqrt{n}$ scaling.

The quantum interferometry is achieved by measuring the intensity difference at the output of interferometer~\cite{H. S. Eisenberg,Itai A,K. J. Resch,D. Leibfried}.
Typically, the sensitivity enhancements in different types of interferometric schemes have been proposed in many setups, such as Sagnac, Mach-Zehnder, Fabry-P\'{e}rot and SU(1,1) interferometers~\cite{K. X. Sun,A. N. Boto,S. J. Bentley,S. H. Tan,V. Giovannetti,M. Tsang,S. Lloyd,G. Khoury,S. D. Huver,B. Yurke}.
Nowadays, instead of exploring quantum metrology in the interferometer, quantum sensors composed by for example quantum dot and cavity QED system have made great achievements~\cite{M. Kasevich,C. E. Wieman,Xiaoming,Fang,FFujimoto,71,72,MW1,LR,61,62}. As one kind of the simple two-mode resonant cavities, the whispering-gallery-mode (WGM) resonator has become a versatile platform for measuring  the angular velocity based on Sagnac effect~\cite{Karl,CDV}.

The WGM supports the clockwise and counterclockwise propagating optical modes, and the effective coupling between the two optical modes can be induced by coupling to a large detuned two-level atom. Adiabatically eliminating the degree of freedom of the atom, we construct an effective interferometer with SU$(2)$ Lie algebra structure in this paper. It is thus similar to the Mach-Zehnder interferometer and supplies us a way to perform the rotating sensing. The underlying physics is to transform the information about the angular velocity to the phase difference of the two optical modes. The study about the quantum Fisher information (QFI) shows that the effective inter-mode coupling, which encodes the information of rotation into both of the amplitudes and the phases of the wave function, plays a decisive role in achieving the Heisenberg limit for the estimation of angular velocity. Moreover, we find that even a weak non-linearity in the system will further enhance the quantum metrology and defeat the Heisenberg limit by achieving a $1/n^3$ scaling for the QFI. We explain the enhancement by the compressibility of state distribution during the time evolution.

The rest of the paper is  organized as follows. In Sec.~\ref{Model}, we review quantum interferometry and present a feasible experimental scheme in the
WGM optical  microresonator. In Sec.~\ref{Nonlinear effect}, we discuss the effect of the nonlinearity in the system for enhancing the quantum metrology. In Sec.~\ref{con}, we give a short summary. In the appendix,  we present some detailed calculations.

\section{Model and Hamiltonian}
\label{Model}

\subsection{Mach-Zehnder Interferometer}
 \begin{figure}[tbp]
\centering
\includegraphics[width=8cm]{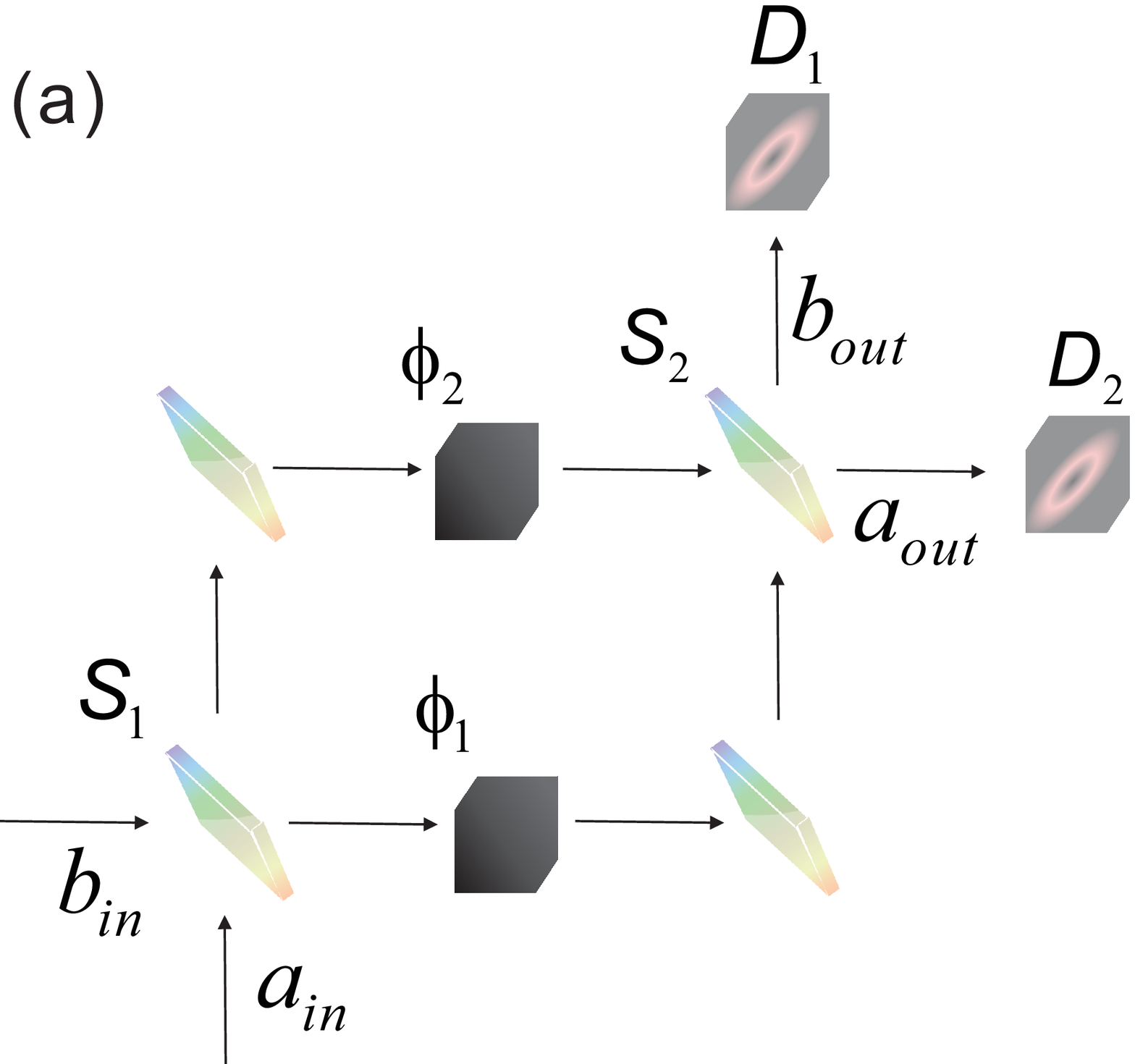}\nonumber\\
\includegraphics[width=8cm]{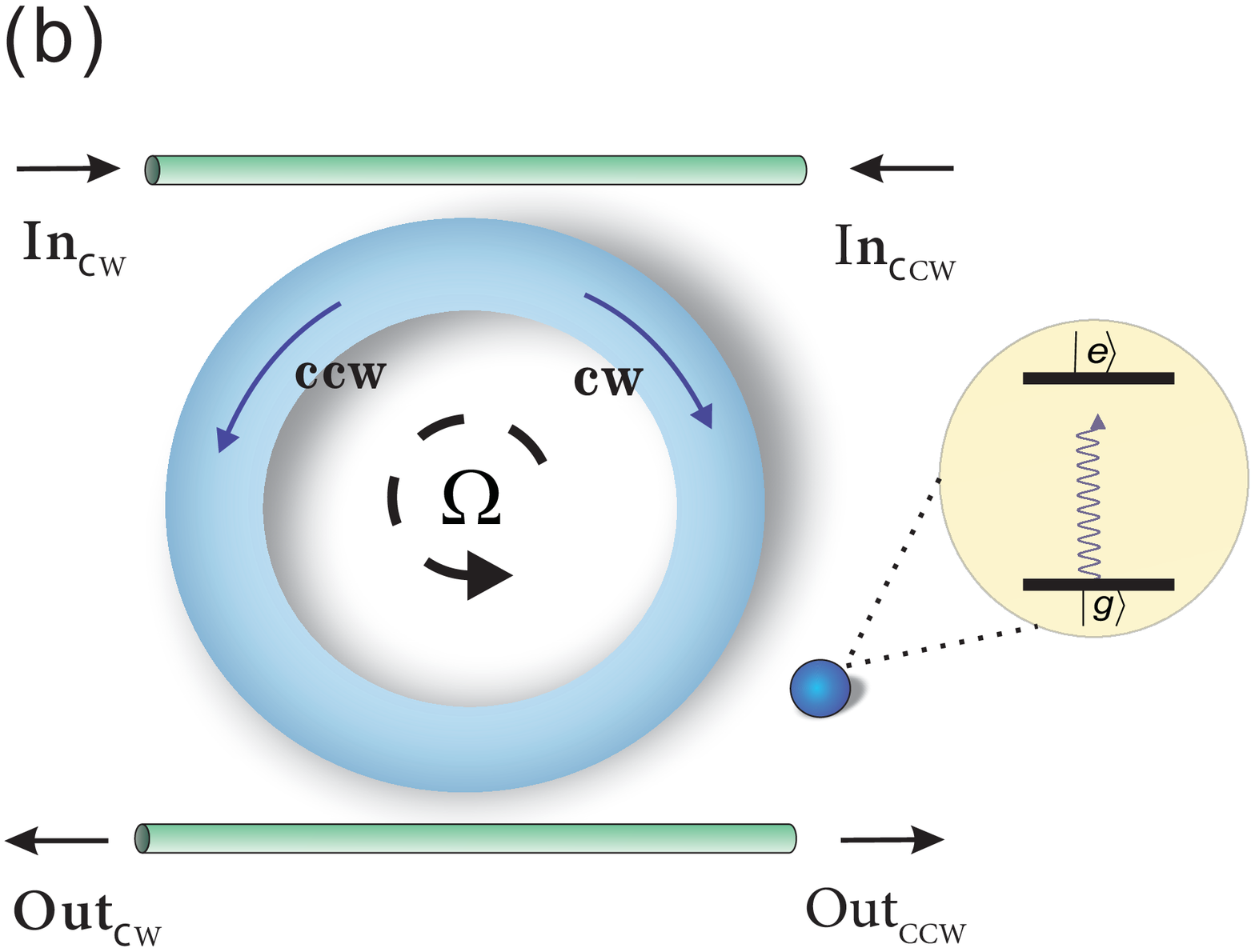}\nonumber\\
\caption{(a) Schematic diagram of the Mach-Zehnder interferometer. It consists of two $50-50$ beam splitters: $S_1$ and $S_2$, the relative phase shift device $\phi=\phi_2-\phi_1$ and the photodetectors: $D_1$ and $D_2$.
(b) Sketch of a rotating whispering-gallery-mode optical resonator, which couples to a large detuned two-level atom. }
\label{mz-interferometer}
\end{figure}
Let us first review the Mach-Zehnder interferometer, which is sketched in Fig.~\ref{mz-interferometer}(a).  It consists of two $50-50$ beam splitters: $S_1$ and $S_2$, the relative phase shift device $\phi=\phi_2-\phi_1$ and the photodetectors: $D_1$ and $D_2$. It is convenient to introduce the Schwinger representation for a two-mode quantized light field, that is
  \begin{eqnarray}
J_{x}&=&\frac{1}{2}(ab^{\dagger}+a^{\dagger}b),\nonumber\\
J_{y}&=&\frac{1}{2i}(ab^{\dagger}-a^{\dagger}b),\nonumber\\
J_{z}&=&\frac{1}{2}(a^{\dagger}a-b^{\dagger}b),\nonumber\\
J^2&=&J^2_x+J^2_y+J^2_z,
\label{Schwinger}
\end{eqnarray}
and
  \begin{eqnarray}
N&=&a^{\dagger}a+b^{\dagger}b.
\label{N}
\end{eqnarray}
Here $a$ and $b$ represent the annihilation operation of two beams
and the commutation relations satisfy the Lie algebra of SU(2):
  \begin{eqnarray}
[J_{x},J_{y}]=iJ_{z},
[J_{y},J_{z}]=iJ_{x},
[J_{z},J_{x}]=iJ_{y}.
\label{commutation}
\end{eqnarray}
It then yields
  \begin{eqnarray}
J_z|j,m\rangle&=&m|j,m\rangle,\nonumber\\
J^2|j,m\rangle&=&j(j+1)|j,m\rangle,
\label{STATE}
\end{eqnarray}
and
  \begin{eqnarray}
  N|j,m\rangle&=&n|j,m\rangle,
\label{NSTATE}
\end{eqnarray}
where $n=2j$ is the total number of photons and $m$ is the difference of photon number between the two ports.

For the sake of convenience, here and after, we set the entangled initial state as $|in\rangle=(|j,0\rangle+|j,1\rangle)/\sqrt{2}$.
In the Schr\"{o}dinger picture, the devices $S_1$, $S_2$ and $\phi=\phi_2-\phi_1$ will lead to transformations $\exp(-i\pi J_x/2)$,  $\exp(i\pi J_x/2 )$ and $\exp(-i\phi J_z)$, respectively. The final state $|out\rangle$ is thus
  \begin{eqnarray}
|out\rangle&=&U(\phi)|in\rangle
=e^{i\frac{\pi}{2}J_x} e^{-i\phi J_z} e^{-i\frac{\pi}{2}J_x}|in\rangle.
\label{out}
\end{eqnarray}

As an interferometer, the phase difference $\phi$ is the parameter to be estimated.
In the field of quantum metrology, the QFI is a central quantity, giving a theoretically achievable limit on the precision for an unknown estimated parameter $\phi$. Considering the parameter $\phi$ as a random variable the mean square fluctuation of $\phi$ is defined as $\Delta \phi=\sqrt{\bar{\phi^2}-\bar\phi^2}$.   According to the quantum Cram\'{e}r-Rao inequality, $\Delta \phi$ is bounded by~\cite{JA,RR,SL1,SL2}
  \begin{eqnarray}
\Delta \phi\geq \frac{1}{\sqrt{\nu\mathcal {F}_{\phi}}},
\label{Cramer-Rao}
\end{eqnarray}
where $\nu$ is the times of the independent measurements and $\mathcal {F}_{\phi}$ is the QFI with respective to
$\phi$. For a general quantum pure state $|\psi\rangle$,  the QFI is given by
  \begin{eqnarray}
\mathcal {F}_{\phi}=4(\langle \partial_\phi\psi|\partial_\phi\psi\rangle-|\langle \psi|\partial_\phi\psi\rangle|^2).
\label{Cramer-Rao}
\end{eqnarray}
Introducing a Hermitian operator
 \begin{equation}
 \mathcal {H}=-iU^{\dagger}(\partial U/\partial \phi)
 \label{huaH1}
  \end{equation}
  ($U$ is the evolution operation), the QFI can be reduced to
  \begin{eqnarray}
\mathcal {F}_{\phi}=4\langle \Delta^2 \mathcal {H}\rangle,
\label{Cramer-Rao}
\end{eqnarray}
 where $\Delta^2 \mathcal{H}= \langle \mathcal {H}^2\rangle-\langle \mathcal {H}\rangle^2$.

 Then we have
  \begin{eqnarray}
\mathcal {F}_{\phi}=2j(j+1)-1,
\label{QFI1}
\end{eqnarray}
so that $\Delta \phi\simeq \sqrt{2}/n$, that is, the fluctuation of $\phi$ is proportional to $1/n$, which refers as the Heisenberg limit. Furthermore,  we note that the QFI will decline dramatically to $1$ without the $50-50$ beam splitters. In fact, the $50-50$ beam splitters, whose roles are described by $e^{\pm i(\pi/2)J_x}$ in Eq.~(\ref{out}), will induce the effective coupling between $a$ and $b$ modes and therefore improve measurement accuracy.

\subsection{Cavity QED setup}
\label{Cavity QED setup}
In the above subsection, we have exhibited the effect of the  Mach-Zehnder interferometer in regard to achieving the Heisenberg limit for the phase estimating.
In this subsection, mimicking the  parametric process in the Mach-Zehnder interferometer, we design the quantum  sensors for rotation in a WGM optical microresonator. The central idea is that the information of the angular velocity of the rotation is transferred to the effective phase between the two ports.

As one of the most simple two-mode resonant cavity, the WGM optical  microresonator shows excellent performance in measuring the angular velocity~\cite{WW,JL}, and the similar Sagnac effect based rotation rate sensitivity is reported to be with sub-prad/s~\cite{CDV}, which is even beyond earth rotation rate ($7.292\times10^{-5}$\,rad/s). We now apply the optical microresonator which couples to a two-level large detuned atom to perform a quantum sensing as illustrated in Fig.~\ref{mz-interferometer}(b).
Here, the atom  with energy separation $\omega_{a}$ between the ground state $|g\rangle$ and excited state $|e\rangle$ is placed near the resonator with an optical resonance frequency $\omega_{l}$. The optical microresonator supports two resonant modes, which are propagated clockwisely (CW) and counterclockwisely (CCW) and as shown below, the two-level atom will induce a weak effective interaction between the two modes.
Furthermore, we introduce two parallel waveguides, which couple to the microresonator. The waveguides can be applied to prepare the input state and perform the measurement on the output photons of the CW and CCW modes as shown in Fig.~\ref{mz-interferometer} (b)\cite{book}.

We consider a gyro setup that the waveguides are stationary while the microresonator is rotated with angular velocity $\Omega$.  Such spinning resonator has been realized experimentally and demonstrate the photonic non-reciprocal transmission~\cite{huijing}. Thanks to the rotation, The optical resonance frequency will be modified $\omega_{l}\rightarrow\omega_{l}\pm\Delta$ due to Sagnac effect~\cite{GB}, where

 \begin{eqnarray}
\Delta=\frac{n_0R\Omega\omega_l}{c}(1-\frac{1}{n_0^2}-\frac{\lambda}{n_0}\frac{dn_0}{d\lambda}).
\label{Delta}
\end{eqnarray}
Here, $n_0$ is the refractive index, $R$ is the radius of resonator and $c$ is the speed of light in vacuum.
$\lambda$ is the wavelength of the probe light and the last term $\lambda dn_0/n_0 d\lambda$ originates from the relativistic effect.
The Hamiltonian of the system  can be written as $H=H_{0}+H_{I}$,
where
\begin{eqnarray}
H_{0}=\sum_{\gamma}\omega_{\gamma}a_{\gamma}^{\dagger}a_{\gamma}+\omega_{a}|e\rangle\langle e|,
\label{H0}
\end{eqnarray}
and
\begin{eqnarray}
H_{I}=\sum_{\gamma}(g_{\gamma}a_{\gamma}|e\rangle\langle g|+\rm{H.c}).
\label{3}
\end{eqnarray}
Here, $\gamma={\rm cw,ccw}$ and $\omega_{\rm cw}=\omega_{l}+\Delta$, $\omega_{\rm ccw}=\omega_{l}-\Delta$. The real $g_{\gamma}$ is the coupling strength between the $\gamma$ mode and the two-level atom. $a_{\gamma}$ and $a^{\dagger}_{\gamma}$ are the annihilation and creation operators of the $\gamma$ mode, respectively.  By use of the Fr\"{o}lich-Nakajima transformation (see Appendix~\ref{A1}) and the Schwinger representation, the approximate effective Hamiltonian of the system can be reduced to
   \begin{eqnarray}
\tilde{H}_{\rm eff}=f(\Delta)J_z+dJ_x,
\label{Hamiltonian}
\end{eqnarray}
where $f=2\Delta$ and $d=2g_{\rm eff}$. Here, the effective coupling strength between the two optical modes $g_{\rm eff}$ is (refer the Appendix A for detailed derivations)
\begin{eqnarray}
g_{\rm eff}=\frac{1}{2}\left(\frac{1}{\Delta_{\rm cw}}
+\frac{1}{\Delta_{\rm ccw}}\right)g_{\rm cw}g_{\rm ccw},
\end{eqnarray}
with $\Delta_\gamma=\omega_a-\omega_\gamma$ ($\gamma={\rm CW,CCW}$). $g_{\rm eff}$ implies that the photon in CW (CCW) mode is virtually absorbed by the atom, and remitted it to the CCW (CW) mode, therefore the two modes couple to each other via a second order process.

Similar to the Mach-Zehnder interferometer, the parametric process is governed by the evolution $U=\exp(-i\tilde{H}_{\rm eff}t)$. We would like to emphasize that the roll of the beam splitters is replaced by the atom, which induces the effective coupling between the two modes, and $d$ in Eq.~(\ref{Hamiltonian}) characterizes the coupling strength.

In order to calculate the QFI  with respective to $\Delta$, we give directly the Hermitian operator in Eq.~(\ref{huaH1}) as~\cite{X. X. Jing,J. Liu}
   \begin{eqnarray}
\mathcal {H}_{\rm eff}=C_xJ_x+C_yJ_y+C_zJ_z,
\label{Hermitian}
\end{eqnarray}
where
   \begin{eqnarray}
C_x&=&\frac{df}{r^3}\frac{\partial f}{\partial \Delta}[\sin(rt)-rt],\nonumber\\
C_y&=&\frac{d}{r^2}\frac{\partial f}{\partial \Delta}[\cos(rt)-1],  \nonumber\\
C_z&=&-\frac{d^2}{r^3}\frac{\partial f}{\partial \Delta}[\sin(rt)+\frac{f^2rt}{d^2}],
\label{Cr}
\end{eqnarray}
and $r=\sqrt{f^2+d^2}$.
Thus
   \begin{eqnarray}
 \mathcal{F}_{\phi}&=&[\frac{n}{2}(\frac{n}{2}+1)-1]C_x^{2}\nonumber\\
&+&2[\frac{n}{2}(\frac{n}{2}+1)-\frac{1}{2}]C_y^{2}+C_z^{2}.
\label{F2}
\end{eqnarray}
 Obviously, we have $ \mathcal{F_\phi}\sim n^2$, which achieves the Heisenberg limit. However, when $d=0$, it will become $ \mathcal{F_\phi}=(\partial f/\partial \Delta) t^2$,  which is independent of $n$ and is smaller than that for $d\neq0$. It indicates that the effective coupling between two modes, which is induced by the large-detuned atom, plays a pivotal role for enhancing quantum metrology and achieving the Heisenberg limit.

{The enhancement of QFI originates from the special encoding scheme for a quantum state. Without inter-mode coupling (i.e., $d=0$), the information about $\Delta$ is only encoded in the phase of the quantum state. However, the atom induced coupling makes not only the phase but also the amplitude contain the information about rotation ($\Delta$). As derived in Appendix~\ref{A2}, we will get an enhancement for QFI.} Specifically, when $d=0$, the dynamical evolution is obtained by $|\psi(t)\rangle=\exp(-i(fJ_z t)|in\rangle=(|j,0\rangle+\exp(-ift)|j,1\rangle)/\sqrt{2}$ and thus $\Delta$ is only encoded in the phase.
However, for $d\neq0$, the evolution will  become complicated: $|\psi(t)\rangle=\exp(-i(fJ_z+dJ_x)t)|in\rangle$.
Under this circumstance, the information of $\Delta$ is not only carried in the phase, but also in the probability amplitude, leading to an enhancement of the quantum metrology.

\section{Nonlinear effect}
 \begin{figure}[tbp]
\centering
\includegraphics[width=8cm]{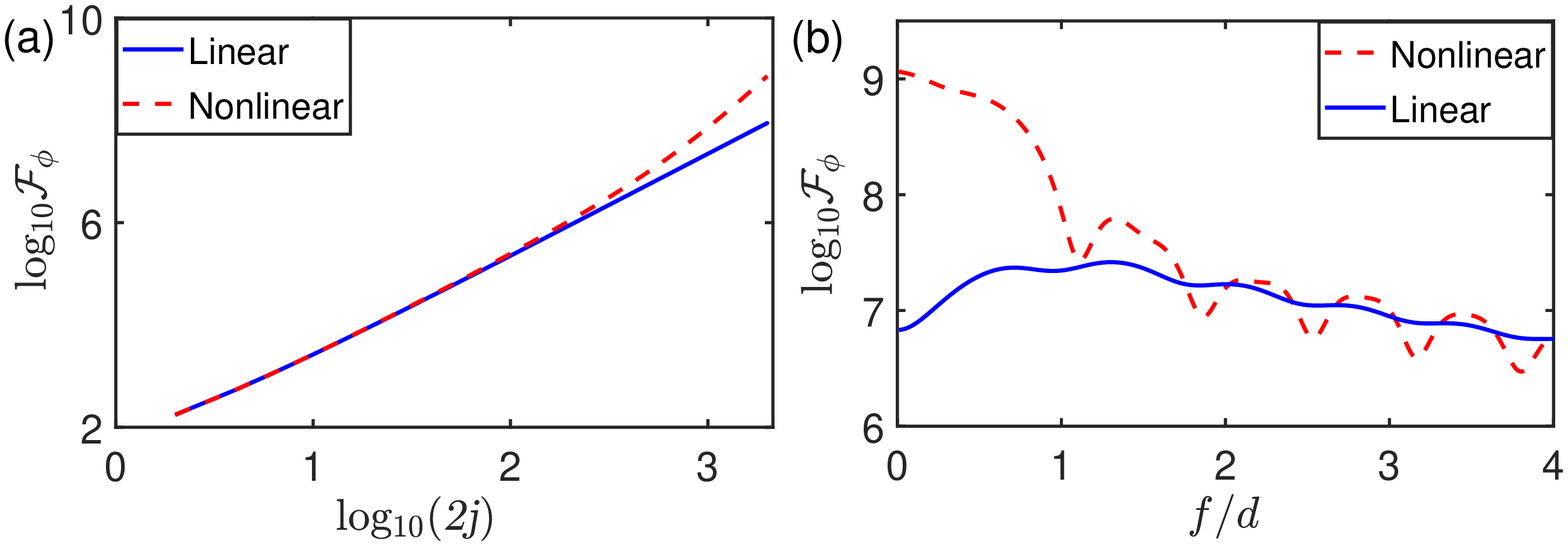}\nonumber\\
\caption{(a) QFI versus the photon number $2j$, $ft=dt=10$.
(b) QFI versus the parameter $f$ for $j=500$, $dt=10$. The parameters are set as $e=0$ for linear scheme and $e=0.01d$ for nonlinear scheme.}
\label{nonlinear1}
\end{figure}
\label{Nonlinear effect}
In the above section, we have outlined that the effective coupling between the CW and CCW  modes, which is induced by the two-level atom, plays a vital role in achieving the Heisenberg limit. In order to further improve the measurement accuracy, we consider an extra
nonlinear term in this section.

For the general nonlinear microcavity system, the Kerr-type is one of the most common form, which is hosted in a cavity that are filled by the atoms with a particular laser-driving four-level structure ~\cite{M. J. Hatmann1,M. J. Hatmann2}.
Combining the above linear model, the current system can be described by the Bose-Hubbard Hamiltonian
 \begin{eqnarray}
H&=&\sum_{\gamma}\{\omega_{\gamma}a_{\gamma}^{\dagger}a_{\gamma}+
U(a_{\gamma}^{\dagger}a_{\gamma}a_{\gamma}^{\dagger}a_{\gamma}
-a_{\gamma}^{\dagger}a_{\gamma})\}\nonumber\\
&+&g_{\rm eff}(a_{\rm cw}^{\dagger}a_{\rm ccw}+a_{\rm ccw}^{\dagger}a_{\rm cw}),
\label{Hamiltonian2}
\end{eqnarray}
where $U$ is on-side interaction strength and $\gamma={\rm cw,ccw}$.
In the Schwinger representation, it can be reduced to
 \begin{eqnarray}
H&=&(\omega_l+U)N+\frac{U}{2}N^2+2UJ_z^2\nonumber\\
&+&2\Delta J_z+2g_{\rm eff}J_x.
\label{Hamiltonian2}
\end{eqnarray}
{For a fixed photon number, the first two terms are constant and  the Hamiltonian is equivalent to  $H=H_0+H_1$,
where $H_0$ is given by Eq.\eqref{Hamiltonian} and }
 \begin{eqnarray}
H_1&=&eJ_z^2
\label{Hamiltonian2}
\end{eqnarray}
with $e=2U$.

 For the nonlinear system, it is complicated to compute the QFI directly.
 However, considering a weak nonlinear effect $e\ll d$, we  keep to the first order of $e$ and the results yield $\mathcal {H}_{\rm non}=\mathcal {H}_0+\mathcal {H}_1$, where $\mathcal {H}_0$ is given by Eq.\eqref{Hermitian} and
   \begin{eqnarray}
\mathcal {H}_1&=&\sum_{\alpha,\beta\neq\alpha}C_{\alpha\beta}\{J_{\alpha},J_{\beta}\}\nonumber\\
&+&C_{xx}(J_{x}^2-J_{y}^2)+C_{yy}(J_{y}^2-J_{z}^2)
\label{CC}
\end{eqnarray}
and $\alpha,\beta=x,y,z$, $\{J_{\alpha},J_{\beta}\}=J_{\alpha}J_{\beta}+J_{\beta}J_{\alpha}$,  $C_{\alpha\beta}$, $C_{xx}$ and $C_{yy}$ are given in Appendix~\ref{A3}.

The analytical results of the QFI are still tedious, so we only give the numerical results here.
The linear and nonlinear contributions lead to a competition of the terms of QFI with different dependence on the total photon number. In Fig.~\ref{nonlinear1}(a), we compare the QFI as functions of the photon number between linear ($e=0$) and nonlinear ($e\neq0$) setups. The difference between red dashed and blue solid lines demonstrates the nonlinear effect.
The results show that the nonlinear effect plays a leading role on QFI when the photon number is large enough. In general, the nonlinear effect is beneficial to break the Heisenberg limit~\cite{L. Pezze,Jose Beltran,A. Luis}. Since the complete expression of QFI [with $\mathcal{H}$ being given by Eq.~(\ref{CC})] is too tedious, in Appendix~\ref{A3} we only give one term in the results by Eq.~(\ref{term1}), which achieves  that scales as $n^3$, being much better than the Heisenberg limit $n^2$  even within a low nonlinear effect. Furthermore, while $d=0$, the QFI will degrade into $\mathcal {F}=(\partial f/\partial \Delta) t^2$, which is consistent with the linear one.
In addition, in Fig.~\ref{nonlinear1}(b), we plot the QFI as a function of the parameter $f$ on a log-log scale for the linear and nonlinear scheme. It can be observed clearly that the nonlinear curve is much larger than the linear one for $f\lesssim d$. However, for $f\gtrsim d$, it becomes choppy in the linear region.

\begin{figure}[tbp]
\centering
\includegraphics[width=8cm]{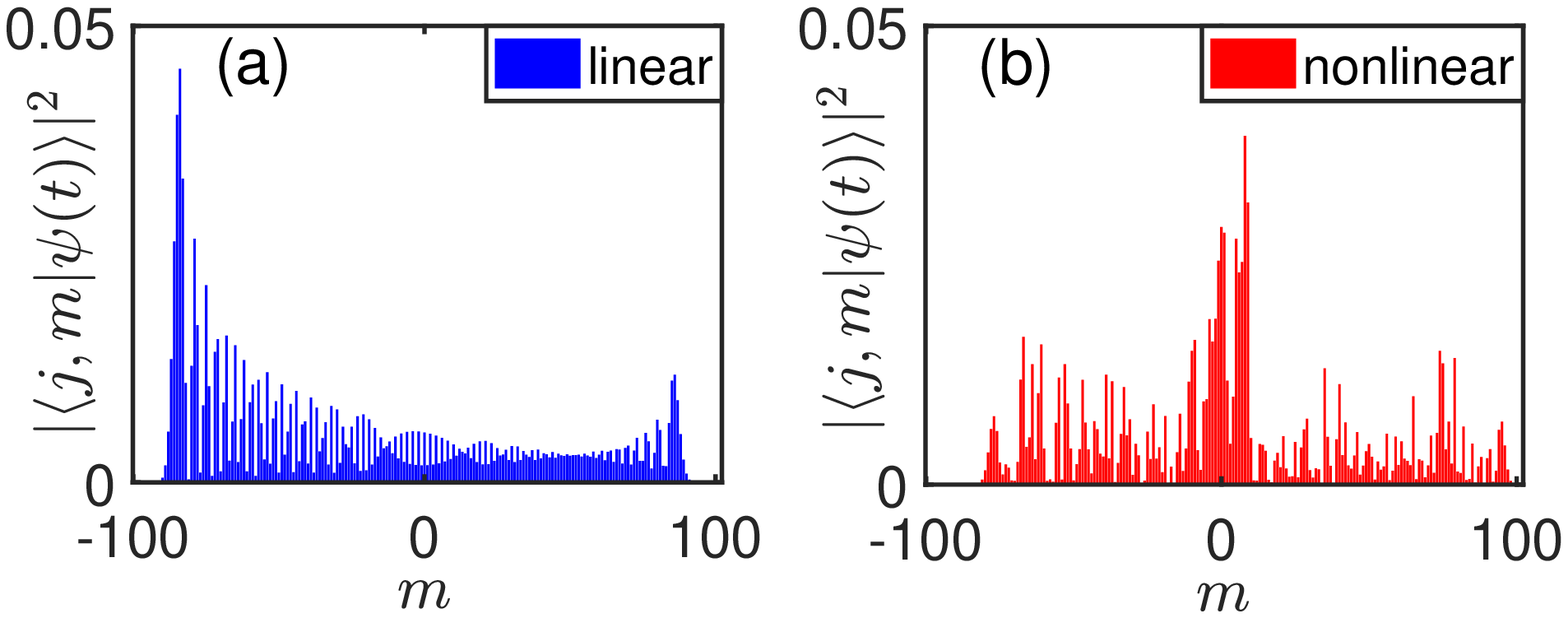}\nonumber\\
\includegraphics[width=8cm]{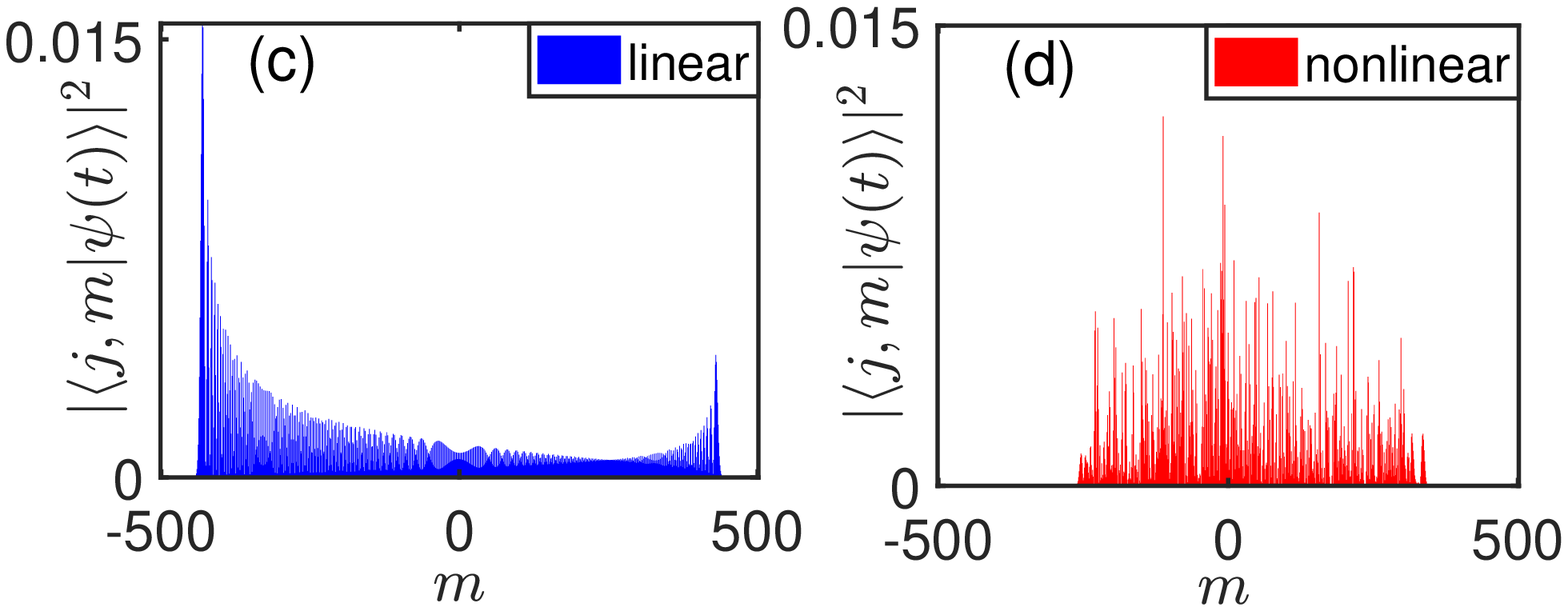}\nonumber\\
\caption{ Distribution of  $|\langle j,m|\psi(t)\rangle|^2$ versus $m$ for linear scheme[(a) and (c)] and nonlinear scheme[(b) and (d)]. For (a) and (b), the parameters are set as $j=100$, $ft=dt=10$, $e=0$ for (a) and $e=0.01d$ for (b). For (c) and (d), the parameters are set as $j=500$, $ft=dt=10$, $e=0$ for (c) and $e=0.01d$ for (d).}
\label{distribution}
\end{figure}

In fact, the distribution of the state has exerted a decisive effect in the above metrology process. As an illustration, we plot the function $|\langle j,m|\psi(t)\rangle|^2$ versus $m$ in Fig.~\ref{distribution}. For $j=100$, in Fig.~\ref{distribution}(a) and (b), we plot $|\langle j,m|\psi(t)\rangle|^2$ for linear and nonlinear scheme, respectively. It shows that the distribution for nonlinear scheme are radically different from linear one, however, their distribution range are approximately the same. For $j=500$, analogously to Fig.~\ref{distribution}(c) and (d), the function $|\langle j,m|\psi(t)\rangle|^2$ versus $m$ are plotted. It is obvious that the distribution for nonlinear scheme [Fig.~\ref{distribution}(d)] are more compressed than the one for linear scheme [Fig.~\ref{distribution}(c)].
{Recall that we have shown in Fig.~\ref{nonlinear1}(a), the linear QFI and the nonlinear QFI are almost unanimous for $j=100$, however, the nonlinear QFI is much larger than the linear one for $j=500$. Therefore, the nonlinearity can induce the compressibility of state distribution, thereby enhancing the QFI of system.}

 \begin{figure}[tbp]
\centering
\includegraphics[width=8cm]{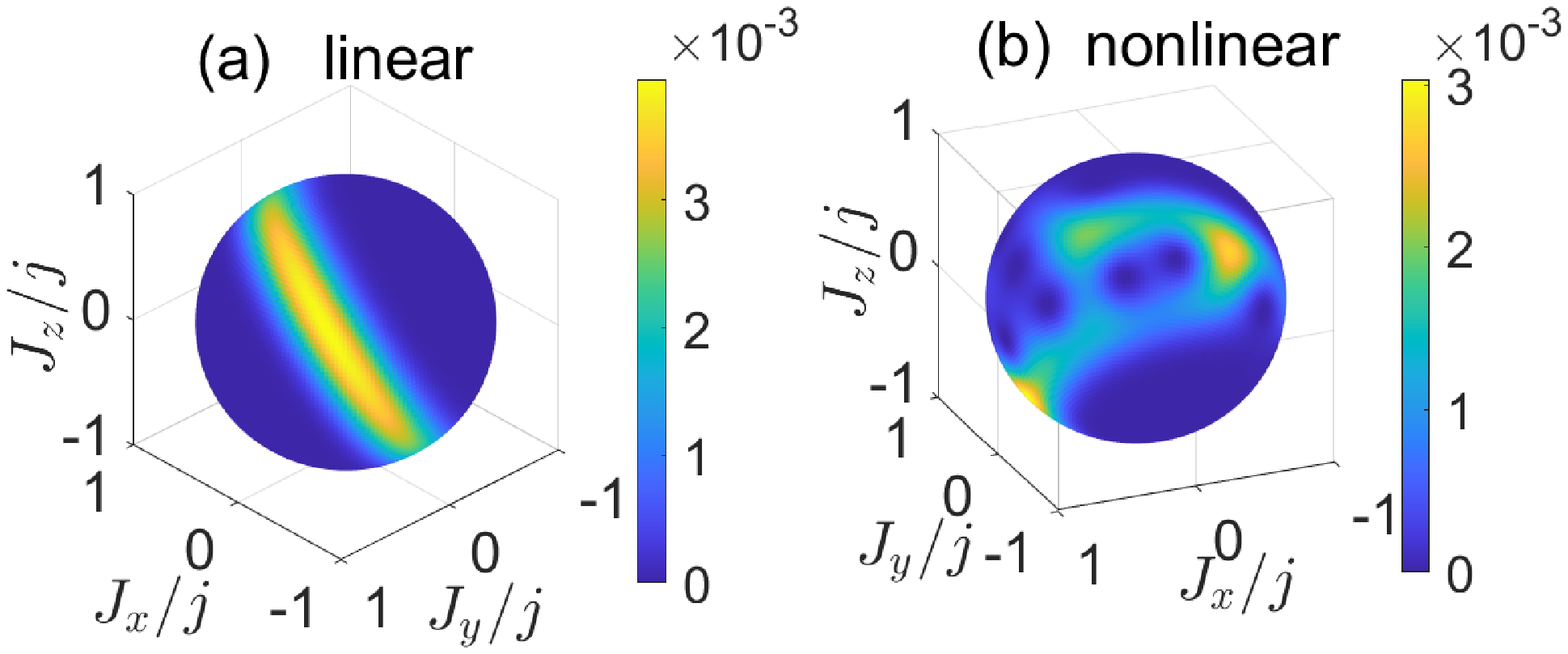}\nonumber\\
\includegraphics[width=8cm]{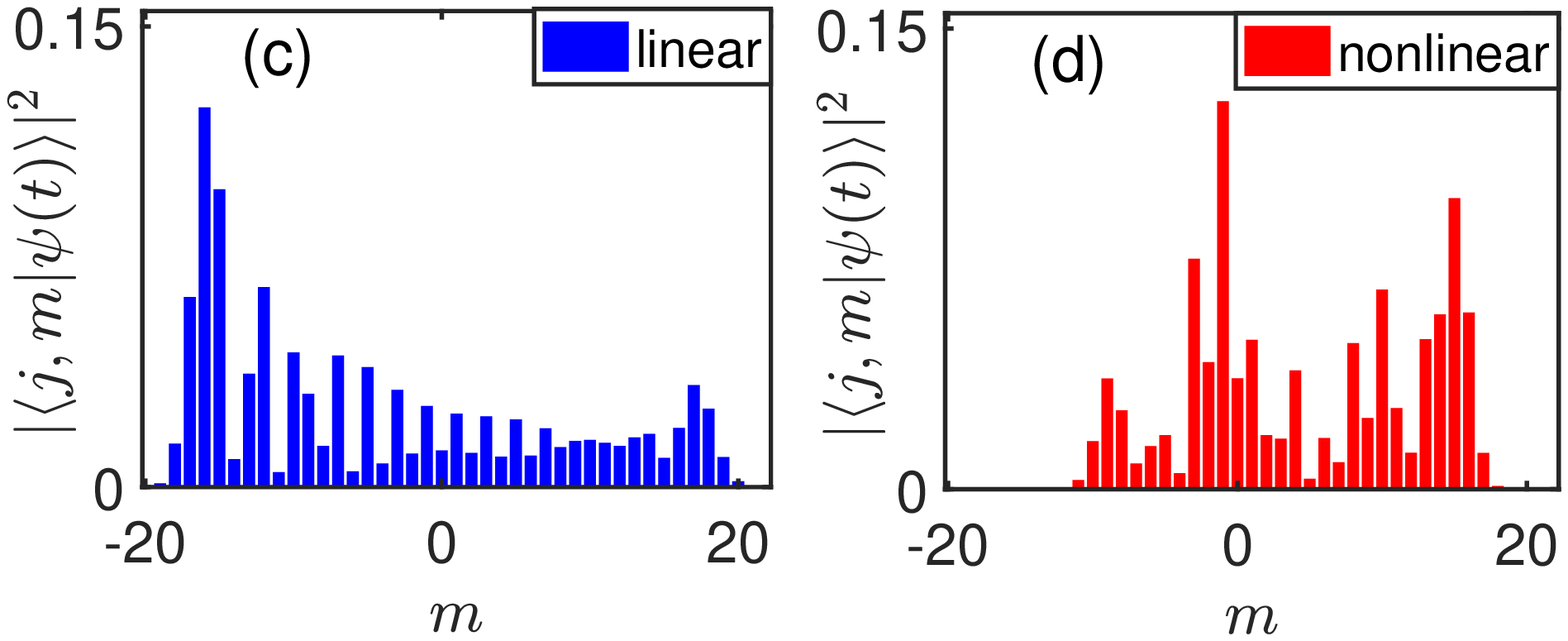}\nonumber\\
\caption{(a) and (b), Q function versus the angles $\theta_0$ and $\phi_0$ for linear scheme and nonlinear scheme, respectively. The parameters are set as $j=20$, $ft=dt=10$, $e=0$ for (a) and $e=0.2d$ for (b).
(c) and (d), distribution of  $|\langle j,m|\psi(t)\rangle|^2$ versus m for linear scheme and nonlinear scheme, respectively.  The parameters are same are set as $j=20$, $ft=dt=10$, $e=0$ for (c) and $e=0.2d$ for (d).}
\label{Qfun}
\end{figure}
The state distribution can be also illustrated  by the Husimi Q function, which represents the anisotropic quasiprobability distribution in a spherical phase space.
The Q function is defined as~\cite{WA,JM}
 \begin{eqnarray}
Q(\theta_0,\phi_0)=\frac{1}{\pi}\langle \theta_0,\phi_0|\rho|\theta_0,\phi_0\rangle,
\label{Q}
\end{eqnarray}
where $|\theta_0,\phi_0\rangle$ is the coherent spin state
 \begin{eqnarray}
|\theta_0,\phi_0\rangle=\exp\{i\theta_0 [J_x\sin(\phi_0)-J_y\cos(\phi_0)]\}|j,-j\rangle\nonumber \\
\label{cherentspin}
\end{eqnarray}
and $\rho$ is the density matrix of the considered system.
In Figs.~\ref{Qfun}(a) and ~\ref{Qfun}(b), we plot Q functions for the linear and nonlinear scheme, respectively.
Compared with the linear scheme, the central area for the nonlinear one [here the nonlinearity strength is $20$ times larger than that in Figs.~\ref{distribution} (b) and (d)] becomes much smaller. Combining the scale of the color bar, we can see that the Q function distribution becomes more uniform for the nonlinear setup.
Meanwhile, we plot $|\langle j,m|\psi(t)\rangle|^2$ versus $m$ under the same condition (the few number of photons and the relatively big nonlinear effect) in Fig.~\ref{Qfun}(c) and ~\ref{Qfun}(d). In this case, the compressibility of state distribution for nonlinear scheme  are revealed again.

\section{conclusion}
\label{con}

In conclusion,  with reference to the traditional SU(2) interferometer model, we implement a similar algebraic setup for the rotation sensing in a microcavity.
In our scheme, we employ a detuning two-level atom to create the effective coupling between two optical modes, thanks to which the initial state is extended to the entire Hilbert space during the time evolution. We find that the accuracy of parameter estimation can be enhanced dramatically by the coupling, via encoding the estimated angular velocity into both of the amplitudes and the phase of the wave function.  Moreover, we study the nonlinear system which can be described by the Bose-Hubbard model and find that the accuracy of parameter estimation can even break the Heisenberg  limit with the large photon number. This enhancement is associated with the compressibility of state distribution.

Our study suggests some viable strategies that may be used to benefit the enhancement of the rotating sensing which include as following: use a large-detuned atom to induce the coupling between two optical modes; introduce some nonlinear interaction in the system; rotating the system with a certain angular velocity.   We hope that our metrology scheme with the assistance of detuning particle can be useful for the designing of quantum gyroscope based on Sagnac effect.

\begin{acknowledgments}
We thank Prof. X.-M. Lu for useful discussions. This work is supported by National Key R$\&$D Program of China (No. 2021YFE0193500), by  National Natural Science Foundation of China
(Grant No.~11875011, No.~12047566, No.~11875231 and No.~11935012.).
\end{acknowledgments}
\appendix
\label{app}

\addcontentsline{toc}{section}{Appendices}\markboth{APPENDICES}{}
\begin{subappendices}
\section{Fr\"{o}lich-Nakajima transformation}
\label{A1}
For our model of the WGM optical microresonator which couples to a two-level detuning atom,
the Hamiltonian can be described by Eq.~\eqref{H0} and \eqref{3} in the main text.
The rotating-wave approximation demands that the coupling strength and detuning satisfy, respectively, $g_{\gamma}\ll\{\omega_{\gamma},\omega_{a}\}$ and $|\Delta_{\gamma}|\ll\{\omega_{\gamma},\omega_{a}\}$, where $\Delta_{\gamma}=\omega_{a}-\omega_{\gamma}$.

To proceed, we assume $g_{\gamma}\ll\Delta_{\gamma}$, the effective coupling between two cavity modes are obtained by the Fr\"{o}lich-Nakajima transformation~\cite{H. B. Zhu,Y. Li C. Bruder,M. Boissonneault}, which is widely used in condensed-matter physics and quantum optics. By eliminating the degree of freedom of the atom, a weak coupling will be established between the two modes. In what follows, we will give the detailed derivations for the transformation.

At first,  we introduce a unitary transformation $\tilde {H}=\exp(-\lambda S)H\exp(\lambda S)$, which can be expanded via Taylor expansions:
\begin{eqnarray}
\tilde {H}& \approx &
H_{0}+\lambda(H_{I}+[H_{0},S])\nonumber\\
&+&\lambda^{2}([H_{I},S]+[S,[S,H_{0}]])+O(\lambda),\nonumber\\
\label{4}
\end{eqnarray}
where $\lambda$ is introduced to mark the order of perturbation and would be set to 1 after all calculations. $S$ is an  anti-Hermitian operator.
Then, setting the first-order perturbation term $H_{I}+[H_{0},S]=0$, we obtain
\begin{eqnarray}
S=g_{\rm cw}a_{\rm cw}|e\rangle\langle g|+g_{\rm ccw}a_{\rm ccw}|e\rangle\langle g|-\rm{H.c}.
\label{5}
\end{eqnarray}
Considering the dispersive  interaction between the atom and resonator modes, we approximate that the atom prepared in initial state $|g\rangle$ will always be in the ground state $|g\rangle$. Neglecting the high-frequency terms, the effective Hamiltonian satisfies
\begin{eqnarray}
\tilde {H}&=&H_{\rm eff}\otimes|g\rangle\langle g|.
\label{6}
\end{eqnarray}
At last, up to second order interactions, our effective Hamiltonian is obtained as $H_{\rm eff}=H_{\rm eff,\omega}+H_{{\rm eff},I}$,
where
\begin{eqnarray}
H_{\rm eff,\omega}=\sum_{\gamma}  \left(\omega_{\gamma+}\frac{|g_{\gamma}|^{2}}{\Delta_{\gamma}}\right)a_{\gamma}^{\dagger}a_{\gamma},
\label{Heffomega}
\end{eqnarray}
and
\begin{eqnarray}
H_{{\rm eff},I}=\sum_{\gamma,\gamma'\neq\gamma} \frac{1}{2}(\frac{1}{\Delta_{\gamma}}+\frac{1}{\Delta_{\gamma'}})
g_{\gamma}g_{\gamma'}a_{\gamma}^{\dagger}a_{\gamma'}.
\label{Heffomega}
\end{eqnarray}

We further define
\begin{eqnarray}
g_{\rm eff}=\frac{1}{2}\left(\frac{1}{\Delta_{\rm cw}}
+\frac{1}{\Delta_{\rm ccw}}\right)g_{\rm cw}g_{\rm ccw},
\end{eqnarray}
in the interaction picture and Schwinger representation, the effective Hamiltonian can  be reduced
 \begin{eqnarray}
\tilde{H}_{\rm eff}&=&\exp(iH_{{\rm eff},\omega}t)H_{{\rm eff},I}\exp(-iH_{{\rm eff},\omega}t)\nonumber\\
&=&2\Delta J_{z}+2g_{\rm eff} J_{x}.
\label{11}
\end{eqnarray}
which is Eq.~(\ref{Hamiltonian}) in the main text.

\begin{figure}
  \centering
  \includegraphics[width=0.8\columnwidth]{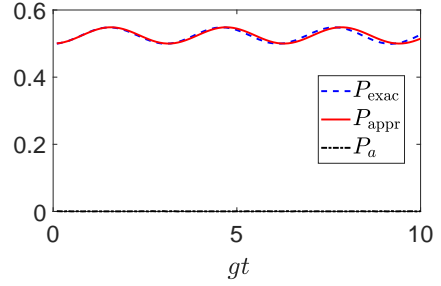}\nonumber\\
  \caption{Dynamics of the system which is governed by the exact and approximate Hamiltonian. The parameters are set as
  $n=20$, $\Delta=g$, $\omega_l=1600g$, $\omega_a=2000g$ and $g_{\rm eff}=g^2/(\omega_a-\omega_l)$. }
\label{prove}
\end{figure}

To verify the above approach, we respectively employ the exact Hamiltonian [see Eq.~\eqref{H0} and \eqref{3}] and the approximate Hamiltonian $\tilde{H}_{\rm eff}$ to illustrate the dynamics of system.
Choosing the initial state as $|in\rangle=(|n,n,g\rangle+|n+1,n-1,g\rangle)/\sqrt{2}$ and $|in\rangle=(|n,n\rangle+|n+1,n-1\rangle)/\sqrt{2}$ for exact and approximate approaches respectively.
We compare the dynamics of of the system in Fig.~\ref{prove}. Here, we plot the dynamics of $P_{\rm exac}=|\langle n,n|\psi_{\rm exac}(t)\rangle|^2$ (the blue dotted line) and the detuning atom $P_{a}=|\langle e|\psi_{\rm exac}(t)\rangle|^2$ (the black dotted line) for the exact solution. Moreover, for the approximate solution, we plot the dynamics of $P_{\rm appr}=|\langle n,n|\psi_{\rm appr}(t)\rangle|^2$ (the red solid line). The agreement between $P_{\rm exac}$ and $P_{\rm appr}$ shows the validity of our approach.  And the fact $P_{a}\approx0$ during the time evolution gives a numerical verification of the Eq.~(\ref{6}).

\section{QFI in different state}
\label{A2}

{In the main text, we have mentioned that the QFI is enhanced  by  encoding the parameter $\Delta$ into both of the amplitudes and the phase of the wave function.} We will give more details in this appendix.
To this end, we now define a general pure state
\begin{eqnarray}
|\psi\rangle&=&\sum_n a_n \exp\left(i\varphi_n\right) |\psi_n\rangle.
\label{stateA1}
\end{eqnarray}
where $\varphi_n$ and $a_n$ are the real phase and the amplitude of the state $|\psi_n\rangle$, respectively, and the normalization condition demands $\sum_n a_n^2=1$.
To analyze the QFI in different state, we consider the following two situations.

First, we consider the case that only the phases are dependent on the estimated parameter, that is $\varphi_n=\varphi_n(\Delta)$. Then the QFI is
\begin{eqnarray}
\mathcal {F}_{1,\Delta}=4\left[\sum_na_n^2\left(\frac{\partial \varphi_n}{\partial \Delta}\right)^2-\mid\sum_n a_n^2\frac{\partial \varphi_n}{\partial \Delta}\mid^2\right].
\label{QFIphi1}
\end{eqnarray}

Second, when the  phases and  the probability amplitudes are both related to $\Delta$, the QFI can be obtained by
  \begin{eqnarray}
\mathcal {F}_{2,\Delta}=\mathcal {F}_{1,\Delta}+4\sum_n\left(\frac{\partial a_n}{\partial \Delta}\right)^2.
\label{QFIphi2}
\end{eqnarray}
Here, the second term $4\sum_n\left(\frac{\partial a_n}{\partial \Delta}\right)^2$ is  the contribution from the probability amplitudes with parameter $\Delta$.
Obviously, encoding the information of $\Delta$ into both of the phases and the amplitudes is beneficial for parameter estimation.

\section{QFI for nonlinear effect}
\label{A3}
In the Eq.~\eqref{CC} of the main text, we have obtained the formal solution of the Hermitian operator $\mathcal {H}$. Here, we give the concrete expression and  derivation process through the Baker-Hausdoff formula~\cite{S. S. Pang,J. Liu,X. X. Jing}:
   \begin{eqnarray}
\mathcal {H}=-t\frac{\partial H}{\partial \Delta}+i\sum_{n=1}^{\infty}\frac{(it)^{n+1}}{(n+1)!}H^{\times n}\frac{\partial H}{\partial \Delta}
\label{huaH}
\end{eqnarray}
where the super operator $H^{\times n}$ denotes a $n$th-order nested commutator operation, $H^{\times}(\cdot)=[H,\cdot]$. Then we have
 \begin{eqnarray}
\mathcal {H}&=&C_xJ_x+C_yJ_y+C_zJ_z+C_{xx}(J_{x}^2-J_{y}^2)\nonumber \\&&+C_{yy}(J_{y}^2-J_{z}^2)+C_{xy}(J_{x}J_{y}+J_{y}J_{x})\nonumber\\&&
+C_{yz}(J_{y}J_{z}+J_{z}J_{y})+C_{zx}(J_{z}J_{x}+J_{x}J_{z}),
\label{MH}
\end{eqnarray}
where
   \begin{eqnarray}
C_{xy}&=&-\frac{e}{6f^2}\frac{\partial f}{\partial \Delta}\{\frac{2A_1}{r^2}[\cos(rt)-1]\nonumber\\
&-&\frac{A_1-B_1/\eta}{\Lambda_1^2}[\cos(\Lambda_1t)-1]\nonumber\\
&-&\frac{A_1+B_1/\eta}{\Lambda_2^2}[\cos(\Lambda_2t)-1]\},
\label{CC1}
\end{eqnarray}
   \begin{eqnarray}
C_{yz}&=&\frac{e}{3fd}\frac{\partial f}{\partial \Delta}\{\frac{A_2}{r^2}[\cos(rt)-1]\nonumber\\
&-&\frac{A_2+B_2/\eta}{2\Lambda_1^3}[\cos(\Lambda_1t)-1]\nonumber\\
&-&\frac{A_2-B_2/\eta}{2\Lambda_2^3}[\cos(\Lambda_2t)-1]\},
\label{CC2}
\end{eqnarray}
   \begin{eqnarray}
C_{zx}&=&\frac{e}{3f^2d}\frac{\partial f}{\partial \Delta}\{\frac{A_3+f^2d^2}{r^3}[\sin(rt)-rt]\nonumber\\
&-&\frac{A_3+B_3/\eta}{2\Lambda_1^3}[\sin(\Lambda_1t)-\Lambda_1t]\nonumber\\
&-&\frac{A_3-B_3/\eta}{2\Lambda_2^3}[\sin(\Lambda_2t)-\Lambda_2t]\},
\label{CC3}
\end{eqnarray}
   \begin{eqnarray}
C_{xx}&=&-\frac{e}{6f}\frac{\partial f}{\partial \Delta}\{\frac{2A_1}{r^3}[\sin(rt)-rt]\nonumber\\
&-&\frac{A_1-B_1/\eta}{\Lambda_1^3}[\sin(\Lambda_1t)-\Lambda_1t]\nonumber\\
&-&\frac{A_1+B_1/\eta}{\Lambda_2^3}[\sin(\Lambda_2t)-\Lambda_2t]\},
\label{CC4}
\end{eqnarray}
and
\begin{eqnarray}
C_{yy}&=&\frac{e}{3f}\frac{\partial f}{\partial \Delta}\{\frac{2A_2}{r^3}[\sin(rt)-rt]\nonumber\\
&-&\frac{A_2+B_2/\eta}{\Lambda_1^3}[\sin(\Lambda_1t)-\Lambda_1t]\nonumber\\
&-&\frac{A_2-B_2/\eta}{\Lambda_2^3}[\sin(\Lambda_2t)-\Lambda_2t]\}.
\label{CC5}
\end{eqnarray}
where $\Lambda_1=[(3f^2+3d^2-\eta)/2]^{1/2}$, $\Lambda_2=[(3f^2+3d^2+\eta)/2]^{1/2}$ , $\eta=[f^4+d^4+14f^2d^2]^{1/2}$ and

  \begin{eqnarray}
\begin{cases}A_1=d^2-4f^2&B_1=(f^2-d^2)(4f^2+d^2)\\
A_2=f^2-d^2 & B_2=f^4-d^4+6f^2d^2\\
A_3=2f^4+d^4-6f^2d^2&B_3=2f^6+d^6+8f^4d^2+f^2d^4. \nonumber\\
\end{cases}
\end{eqnarray}

The complete expression for the QFI
$\mathcal{F}_\Delta=4(\langle \mathcal{H}^2\rangle-\langle \mathcal{H}\rangle^2)$ [with $\mathcal{H}$ being given by Eq.~(\ref{CC})] possesses $36$ terms, which is too tedious to be given term by term here. However, we note that $\{C_{xx}, C_{yy}, C_{xy}, C_{yz}, C_{zx}\}\propto e^{1}$ and $\{C_x, C_y,C_z\}\propto e^{0}$.  Therefore, in the expression of the QFI, $6$ terms are proportional to $e^0$, $15$ terms are proportional to $e^1$ while the remaining $15$ terms are proportional to $e^2$. The first $6$ terms in the order of $e^0$ are exactly the QFI without the nonlinear interaction, which is given by Eq.~(\ref{F2}) in the main text. The last $15$ terms in the order of $e^2$ can be neglected in the situation of small $e$. Now, we give one typical term which is proportional to $e^1$, for example
\begin{eqnarray}
  &&C_xC_{xx}\{\langle J_x(J_x^2-J_y^2)+(J_x^2-J_y^2)J_x\rangle -2 \langle J_x\rangle \langle J_x^2-J_y^2\rangle\}\nonumber\\
&\propto&\frac{1}{8}C_xC_{xx}[(j-1)(j+2)\sqrt{j(j+1)}].
\label{term1}
\end{eqnarray}
Since $j=n/2$, it shows that we here achieve a $n^3$ scaling for the QFI, which is beyond the Heisenberg limit.

\end{subappendices}

\end{document}